\documentclass[twocolumn,showpacs,amsmatsh,amssymb,aps,prc,groupedaddress,superscriptaddress]{revtex4-1}

\usepackage[caption=false]{subfig}
\usepackage{tabularx}
\usepackage{graphicx}
\graphicspath{{./figs/}}
\usepackage{dcolumn}
\usepackage{bm}
\usepackage{float}
\usepackage[Gray,squaren,thinqspace,thinspace]{SIunits} 
\usepackage{amsmath}
\usepackage{ifthen}
\usepackage{xspace}

\newcommand{\SET}[1]  {\ensuremath{\mathcal{#1}}}

\newcommand{\VEC}[1]  {\ensuremath{\boldsymbol{#1}}}

\begin{document}


\title{$A$--dependence of quasielastic charged--current neutrino--nucleus cross sections}


\author{N.~Van Dessel}
\email{Nils.VanDessel@UGent.be}
\affiliation{Department of Physics and Astronomy,\\
Ghent University,\\ Proeftuinstraat 86,\\ B-9000 Gent, Belgium}
\author{N.~Jachowicz}
\email{Natalie.Jachowicz@UGent.be}
\affiliation{Department of Physics and Astronomy,\\
Ghent University,\\ Proeftuinstraat 86,\\ B-9000 Gent, Belgium}
\author{R.~Gonz\'{a}lez-Jim\'{e}nez}
\affiliation{Department of Physics and Astronomy,\\
Ghent University,\\ Proeftuinstraat 86,\\ B-9000 Gent, Belgium}
\author{V.~Pandey}
\affiliation{Center for Neutrino Physics, \\ Virginia Tech, \\ 
Blacksburg, Virginia 24061, \\
USA}
\author{T.~Van Cuyck}
\affiliation{Department of Physics and Astronomy,\\
Ghent University,\\ Proeftuinstraat 86,\\ B-9000 Gent, Belgium}



\date{\today}

\begin{abstract}
\textbf{Background:} ${}^{12}\mathrm{C}$ has been and is still widely used in neutrino--nucleus scattering and oscillation experiments. More recently, ${}^{40}\mathrm{Ar}$ has emerged as an important nuclear target for current and future experiments. Liquid argon time projection chambers (LArTPCs) possess various advantages in measuring electroweak neutrino-nucleus cross sections. Concurrent theoretical research is an evident necessity. \\

\textbf{Purpose:} ${}^{40}\mathrm{Ar}$ is larger than ${}^{12}\mathrm{C}$, and one expects nuclear effects to play a bigger role in reactions. We present inclusive differential and total cross section results for charged--current neutrino scattering on ${}^{40}\mathrm{Ar}$ and perform a comparison with ${}^{12}\mathrm{C}$, ${}^{16}\mathrm{O}$ and ${}^{56}\mathrm{Fe}$ targets, to find out about the A-dependent behavior of model predictions. \\

\textbf{Method:}
 Our model starts off with a Hartree--Fock description of the nucleus, with the nucleons interacting through a mean field generated by an effective Skyrme force. Long--range correlations are introduced by means of a continuum random phase approximation (CRPA) approach. Further methods to improve the accuracy of model predictions are also incorporated in the calculations. \\
 
 \textbf{Results:} We present calculations for ${}^{12}\mathrm{C}$,  ${}^{16}\mathrm{O}$, ${}^{40}\mathrm{Ar}$ and ${}^{56}\mathrm{Fe}$, showcasing differential cross sections over a broad range of kinematic values in the quasielastic regime. We furthermore show flux--folded results for ${}^{40}\mathrm{Ar}$ and we discuss the differences between nuclear responses. \\
 
 \textbf{Conclusions:} At low incoming energies and forward scattering we identify an enhancement in the ${}^{40}\mathrm{Ar}$ cross section compared to ${}^{12}\mathrm{C}$, as well as in the high $\omega$ (low $T_\mu$) region across the entire studied $E_\nu$ range. The contribution to the folded cross section of the reaction strength at values of $\omega$ lower than 50 MeV for forward scattering is sizeable. 
\end{abstract}

\pacs{25.30.Pt, 13.15.+g, 24.10.Jv, 24.10.Cn, 21.60.Jz}

\maketitle

\section{Introduction}\label{sec:int}
Neutrinos have been a hot topic in physics for quite some time now\cite{Alvarez-Ruso:2017oui,Katori:2016yel}, with areas of research ranging from astrophysical supernovas to neutrino oscillations. A central aspect in all of the experiments that aim at disentangling the properties of these fascinating particles is the interaction probability of neutrinos with atomic nuclei. Due to the fact that these particles only interact weakly, cross sections are inherently very small. This necessitates the use of heavy targets for detection, for which one employs nuclei. One type of experiment where neutrinos collide with nuclei is in accelerator--based ones. In these experiments, the energies of the neutrinos are spread over a large range. The most important reaction channel at intermediate energies is charged--current quasielastic scattering (CCQE), where an incoming neutrino transforms into a charged lepton through the exchange of a $W^+$--boson, which changes a neutron into a proton that gets knocked out of the nucleus. \\
  
In experiments studying neutrino--nucleus interactions, a popular target for neutrinos has traditionally been ${}^{12}\mathrm{C}$, amongst others such as ${}^{1}\mathrm{H}$. Collaborations include MiniBooNE \cite{miniboone}, MINERvA \cite{minerva} and T2K \cite{t2k}, which have all studied differential cross sections for both charged and neutral--current neutrino and antineutrino scattering off various nuclear targets \cite{AguilarArevalo:2010zc, AguilarArevalo:2010cx, AguilarArevalo:2013hm, AguilarArevalo:2013nkf, Fiorentini:2013ezn, Fields:2013zhk, Abe:2013jth, Abe:2014agb, Abe:2015oar, Abe:2014iza}. More recent times have seen the development of several new dedicated experiments such as ArgoNeuT \cite{argoneut}, MicroBooNE \cite{microboone} and DUNE\cite{Dune}, which is planned to start taking data in 2022. These collaborations all make use of Liquid Argon Time Projection Chambers (LArTPCs). These detectors, first proposed by Carlo Rubbia in 1977 \cite{Rubbia:117852}, possess several advantageous qualities \cite{Acciarri:2015hha}. For example, they have unprecedented calorimetric measurement precision, tracking capabilities and hadron detection and identification potential. This allows for very precise cross section measurements with minimal background contamination. More exclusive measurements containing information on the outgoing nucleons are now also possible, with e.g.~ArgoNeuT having reported on the appearance of back--to--back ejected proton pairs, so--called 'hammer events' \cite{Acciarri:2014gev}. Given the use of ${}^{40}\mathrm{Ar}$ as target nucleus, the need for theoretical calculations on neutrino--argon interactions is evident. \\

Several theoretical models exist \cite{Alberico:1981sz, Martini:2009uj, Gil:1997bm, Nieves:2004wx, Nieves:2005rq, Benhar:2005dj, Ivanov:2013bta, Gonzalez-Jimenez:2014eqa, Amaro:2006if, Leitner:2008ue, Meucci:2003cv, Ankowski:2005wi, Martini:2016eec} that aim at calculating CCQE cross section results that predict experimental data. Monte Carlo generators used in data analyses are primarily based on relativistic Fermi gas (RFG) models. While this model is relatively successful at reproducing the main features of CCQE scattering cross sections, it is less suitable to accurately model events of low momentum transfer $q$ (at low $E_\nu$ or for very forward scattering) where nuclear effects play a key role. These manifest themselves, e.g., in the form of low-lying excitations and giant resonances \cite{Pandey:2014tza}. Since neutrino beams are not monochromatic, the measured cross section contains contributions from neutrinos with an energy ranging up to a few GeV. Therefore, it is essential to use models that provide realistic predictions over the whole range of energies. This work aims at calculating CCQE cross sections. Other calculations on ${}^{40}\mathrm{Ar}$ have been presented in Refs.~\cite{Ankowski:2005wi, Butkevich:2012zr,Gallmeister:2016dnq}. \\

The model we employ in this work starts off with a Hartree--Fock (HF) description of the nucleus using an effective Skyrme interaction to describe the nucleon--nucleon interaction, incorporating the effect of long--range correlations through the continuum random phase approximation (CRPA). This approach is suitable for the description of the quasielastic peak and the low--lying collective excitations out of a correlated ground state \cite{Pandey:2014tza, Pandey:2016jju}.  \\

This article is structured as follows. In section ~\ref{sec:for} we briefly summarize the most important aspects of our approach. We then show the results for argon calculations in section \ref{sec:results}, comparing them with results for carbon, oxygen and iron. 

\section{Formalism}\label{sec:for}
In this section, we summarize the key ingredients that make up the framework with which we calculate the cross section. The process we consider is inclusive CCQE neutrino--nucleus scattering:
\begin{equation}\label{eq:reac}
 \nu_\mu + {}^{A}_{Z}\mathrm{X} \rightarrow \mu^- + {}^{A-1}_{Z}\mathrm{X'} + p.
\end{equation} 
The muon neutrino has four--momentum $k^\mu_i = (E_i,\VEC{k}_i)$. Through the emission of a $W^+$--boson, it changes into a muon with four--momentum $k^\mu_f = (E_f,\VEC{k}_f)$. The $W^+$--boson carries four--momentum $q^\mu = (\omega,\VEC{q})$, and strikes the nucleus, which is considered to be at rest in the lab frame.
\begin{align}
 \omega &= E_i-E_f \\
 \VEC{q} &= \VEC{k}_i-\VEC{k}_f.
\end{align}

A single neutron in the nucleus is transformed into a proton, which is ejected from the nucleus. Since we concern ourselves with inclusive calculations, the cross sections have the hadronic part of the final state integrated out. \\

\begin{figure}[ht]
  \centering
  \includegraphics[width=0.85\columnwidth]{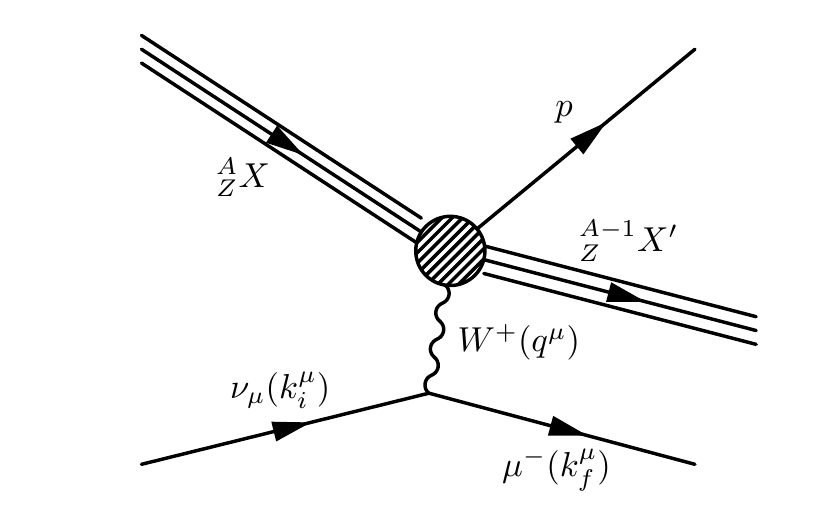}
  \caption{Diagrammatical representation of CCQE neutrino--nucleus scattering}
  \label{fig:srccorr}
\end{figure}

The double differential cross section of this process is expressed as follows
\begin{equation}\label{eq:xsec}
\begin{aligned}
\frac{\mathrm{d}\sigma}{\mathrm{d}T_f\mathrm{d}\Omega_f} =& \left(\frac{G_F \cos{\theta_c}}{2\pi} \right)^2 E_f k_f \zeta^2(Z',E_f,q) \\
&\times \left( v_{CC} W_{CC} + v_{CL} W_{CL} + v_{LL} W_{LL} \right.  \\
& + \left. v_{T} W_{T} \pm v_{T'} W_{T'} \right).
\end{aligned}
\end{equation}
In this expression, $G_F$ is the Fermi constant and $\theta_c$ is the Cabibbo angle. $T_f$ and $\Omega_f$ are the kinetic energy and the solid angle of the outgoing muon, respectively and $\zeta(Z',E_f,q)$ is  the correction factor related to the Coulomb interaction of the outgoing muon with the residual nucleus. The $+(-)$ sign in the final term is applicable to the process of a (anti)neutrino scattering off a nucleus. The $v_i$--factors are functions of the leptonic kinematic variables, whilst the $W_i$--factors are the nuclear response functions, that depend on the transition amplitudes, defined in terms of the nuclear current operator
\begin{equation}
\SET{J}_\lambda^{nucl}(\omega,\VEC{q}) = \langle \Phi_\textrm{f} | \hat{J}_\lambda(\VEC{q}) | \Phi_\textrm{0} \rangle .
\end{equation}
The full expressions of the functions can be found in e.g. \cite{VanCuyck:2016fab}. \\

The starting point of our approach is the description of both initial and final states of the nucleus as a Slater determinant, with single--particle wave functions generated through a HF calculation, using the SkE2 parametrization \cite{Waroquier:1986mj} of the empirical Skyrme contact force \cite{PhysRevC.5.626}, to describe the nucleon--nucleon interaction. The wave functions of the scattered nucleon are calculated in the same mean field as the bound states, thus taking into account the elastic distortion of the nucleon caused by its interaction with the residual nucleus. \\

The calculation of the nuclear response functions is performed within the CRPA \cite{Pandey:2014tza}. Within the RPA, excited states of a many--body system are approximated as coherent superpositions of particle--hole and hole--particle excitations out of a correlated ground state
\begin{equation}
| \Psi^C_{RPA} \rangle = \sum_{C'}\left(X_{C,C'} | p'h'^{-1} \rangle - Y_{C,C'} | h'p'^{-1} \rangle\right).
\end{equation}

In a Green's function formalism, one can then formulate the RPA equations in coordinate space, allowing for an exact treatment of the energy continuum: 

\begin{equation}\label{eq:RPA}
\begin{split}
&\Pi^{(RPA)}(x_1,x_2,E_x) = \Pi^{(0)}(x_1,x_2,E_x) \\ 
& + \frac{1}{\hbar}\int \mathrm{d}x \int \mathrm{d}x' \Pi^{(0)}(x_1,x,E_x)\tilde{V}(x,x')\Pi^{(RPA)}(x',x_2,E_x) ,
\end{split}
\end{equation}
where $x$ is the combined spatial, spin and isospin coordinate, and $E_x$ the excitation energy of the target nucleus. $\Pi^{(RPA)}(x_1,x_2,E_x)$ is the (local) polarization propagator \cite{fetterwalecka1971}, which describes the propagation of particle--hole pairs, and is obtained by adding the iteration of first-order contributions to the bare local polarization propagator $\Pi^{0}(x_1,x_2,E_x)$. $\tilde{V}(x,x')$ is the antisymmetrized residual interaction. We use the same interaction as was used to generate the mean field single particle wave functions, guaranteeing self--consistency in our approach. One can, through analysis of this equation, calculate the nuclear reponse functions within the CRPA through knowledge of the transition amplitudes in the mean field approach \cite{Ryckebusch:1988aa}. We also note that, since we perform calculations on ${}^{40}\mathrm{Ar}$, which is not a closed shell nucleus, we need to take into account partially filled shells. In our calculations, this generalization is readily achieved by including occupation probabilities in the transition amplitudes:
\begin{equation}
\langle ph^{-1} | \hat{O} | \Phi_\textrm{0} \rangle \rightarrow v_h \langle ph^{-1} | \hat{O} | \Phi_\textrm{0} \rangle .
\end{equation}
Herein, $\hat{O}$ represents a general one--body operator and $v_h^2$ is the occupation probability of the shell to which the hole state $h$ belongs. \\

We furthermore introduce several effective schemes to improve the accuracy of the model's predictions, as in Refs.~\cite{Pandey:2015gta,Pandey:2016jju}:

\begin{itemize}
\item The Skyrme force is fitted to the properties of nuclear ground--states and low--lying excitated states. To remedy the interaction's unrealistically high strength at high $Q^2 = -q_\mu q^\mu$ values, we constrain the potential at the strong vertex through a dipole factor \cite{Pandey:2014tza}
 \begin{equation}
V \rightarrow V \frac{1}{\left(1 + \frac{Q^2}{\Lambda^2} \right)^2}.
\end{equation}

The value $\Lambda =$ 455 MeV was optimized in a $\chi^2$ test of the comparison of CRPA cross sections with worldwide A(e,e') experimental data. There, we considered the theory-experiment comparison from low values of omega up to the maximum of the quasielastic peak \cite{Pandey:2014tza}.

\item To take the Coulomb interaction between the outgoing muon and the residual nucleus into account, we use the Modified Effective Momentum Approach (MEMA), which lowers the outgoing muon's momentum (increases for antineutrinos), or equivalently increases (resp.~decreases) the momentum transfer \cite{PhysRevC.57.2004}:
\begin{equation}
q \rightarrow q_{eff} = q \pm 1.5\left(\frac{Z' \alpha \hbar c}{R} \right),
\end{equation}
with $R = 1.24 A^{1/3} \mathrm{fm}$ and $Z'$ the charge of the residual nucleus. The outgoing muon wave function is modified through
\begin{equation}
\Psi_l \rightarrow \Psi_l^{eff} = \zeta(Z',E_f,q) \Psi_l,
\end{equation}
with
\begin{equation}
\zeta(Z',E_f,q) = \sqrt{\frac{q_{eff}E_{eff}}{qE}},
\end{equation}
which modifies the density of final states to be in line with the effective momentum. All of these factors are already included in Equation (\ref{eq:xsec}).
\item We take hadronic relativistic effects into account in an effective way, which can be achieved through following substitution in the nuclear response function \cite{Jeschonnek:1997dm}
\begin{equation}
\lambda \rightarrow \lambda(1 + \lambda),
\end{equation}
with $\lambda = \frac{\omega}{2M_N}$. Furthermore, in the nuclear current, one takes into account relativistic corrections to the usual prescription of the nuclear current, as described in Ref.~\cite{Amaro:2005dn}.

\item As discussed in Ref.~\cite{Pandey:2014tza}, the basic RPA formalism is succesful in predicting the position and strength of giant resonances, but not the shape: the height of the peak is overestimated and the width is underestimated. We use a phenomenological approach to deal with this, by folding the cross section in the following manner
\begin{align}
&\frac{\mathrm{d}\sigma '}{\mathrm{d}\omega\mathrm{d}\Omega_f} (E_\nu, \omega) \\
&= \int \mathrm{d}\omega' L(\omega,\omega') \frac{\mathrm{d}\sigma}{\mathrm{d}\omega'\mathrm{d}\Omega_f}(E_\nu, \omega'),
\end{align}
with
\begin{equation}
L(\omega,\omega')=\frac{1}{2\pi}\left[ \frac{\Gamma}{(\omega-\omega')^2 + (\Gamma/2)^2} \right],
\end{equation}
and $\Gamma\ =$ 3 MeV.
\end{itemize}


The CRPA framework summarized above has seen several successful applications in the description of electroweak probing of nuclei \cite{Ryckebusch:1988aa, Ryckebusch:1989nn, Jachowicz:1998fn, Jachowicz:2002rr, Jachowicz:2004we, Jachowicz:2006xx, Jachowicz:2008kx}. Recent results include a detailed study of the model predictions for inclusive QE electron scattering $(e,e')$ off various nuclear targets such as ${^{12}\textnormal{C}}$, ${^{16}\textnormal{O}}$ and ${^{40}\textnormal{Ca}}$,  alongside comparison with data \cite{Pandey:2014tza}. Furthermore, the application of the CRPA approach to CCQE (anti)neutrino scattering was previously studied for ${^{12}\textnormal{C}}$ nuclei as well, including comparison with MiniBooNE and T2K data \cite{Pandey:2013cca, Pandey:2014tza, Pandey:2016jju}. From these studies we have learned that our approach successfully predicts nuclear excitations at small energy transfer $\omega < 50$ MeV and momentum transfer $q  < 300 $ MeV, that low-lying excitations below the QE peak dominate cross sections at low incoming neutrino--energies $E_\nu$ around a few 100 MeV's, and most importantly that low--lying nuclear excitations are predicted to contribute sizeably to flux--integrated cross sections at forward lepton scattering angles. In forward scattering bins, low--energy nuclear excitations can account for a sizeable fraction of the flux--folded cross section \cite{Pandey:2016jju}.

\section{Results}\label{sec:results}

\begin{figure}
   \centering
   \includegraphics[width=0.6\columnwidth]{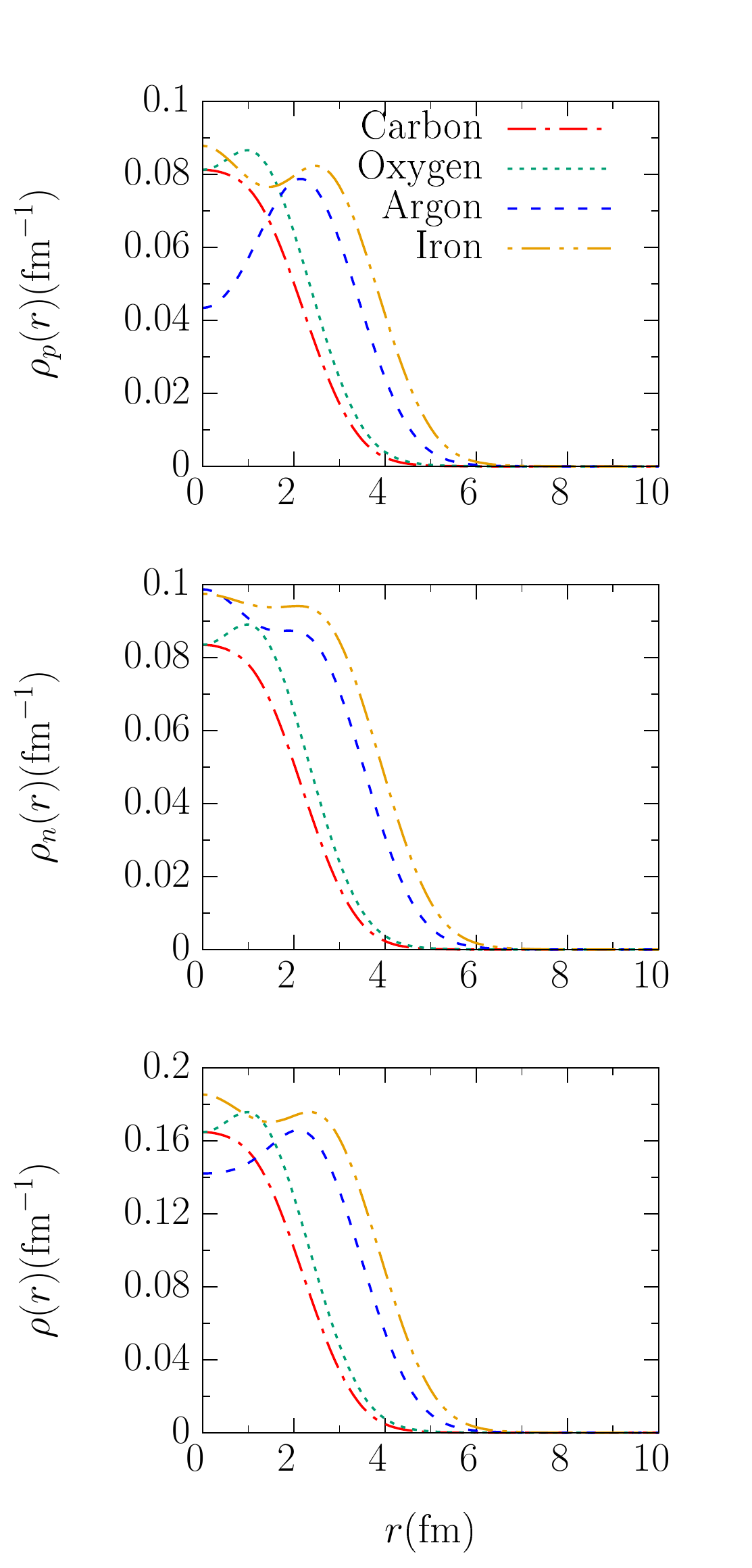}
   \caption{From top to bottom: proton, neutron and nucleon density profiles ($\rho_p, \rho_n, \rho$) for relevant nuclei as yielded by the SkE2 HF calculations.}
   \label{fig:density}
\end{figure}

In our approach, the different structure of various nuclei is taken into account through the use of nuclear wave functions. We describe the nucleus initially as a Slater determinant. Through use of the ansatz $ \psi_{njlm_j}(x) = R_{nlj}(r)\mathcal{Y}_{ljm_j}(\Omega,\sigma)$ in combination with a Skyrme interaction, one can implement a HF calculation which yields the radial single particle wave functions $R_{nlj}(r)$ for the different proton (p) and neutron (n) shells. The shells are then filled up according to the Pauli principle. With these wave functions one can construct the nuclear density profiles. These are pictured in Fig.~\ref{fig:density}, and are normalized such that, e.g.:
\begin{equation}
4\pi\int_0^{+\infty} \mathrm{d}r r^2 \rho_p(r) = Z.
\end{equation}
The neutron and nucleon densities are normalized in a similar fashion to $N$ and $A$, respectively. One notices the increasing size of the nucleus, as well as the lower central values of the proton densities due to Coulomb repulsion for larger Z. \\

\begin{figure*}
   \centering
   \includegraphics[width=1.0\textwidth]{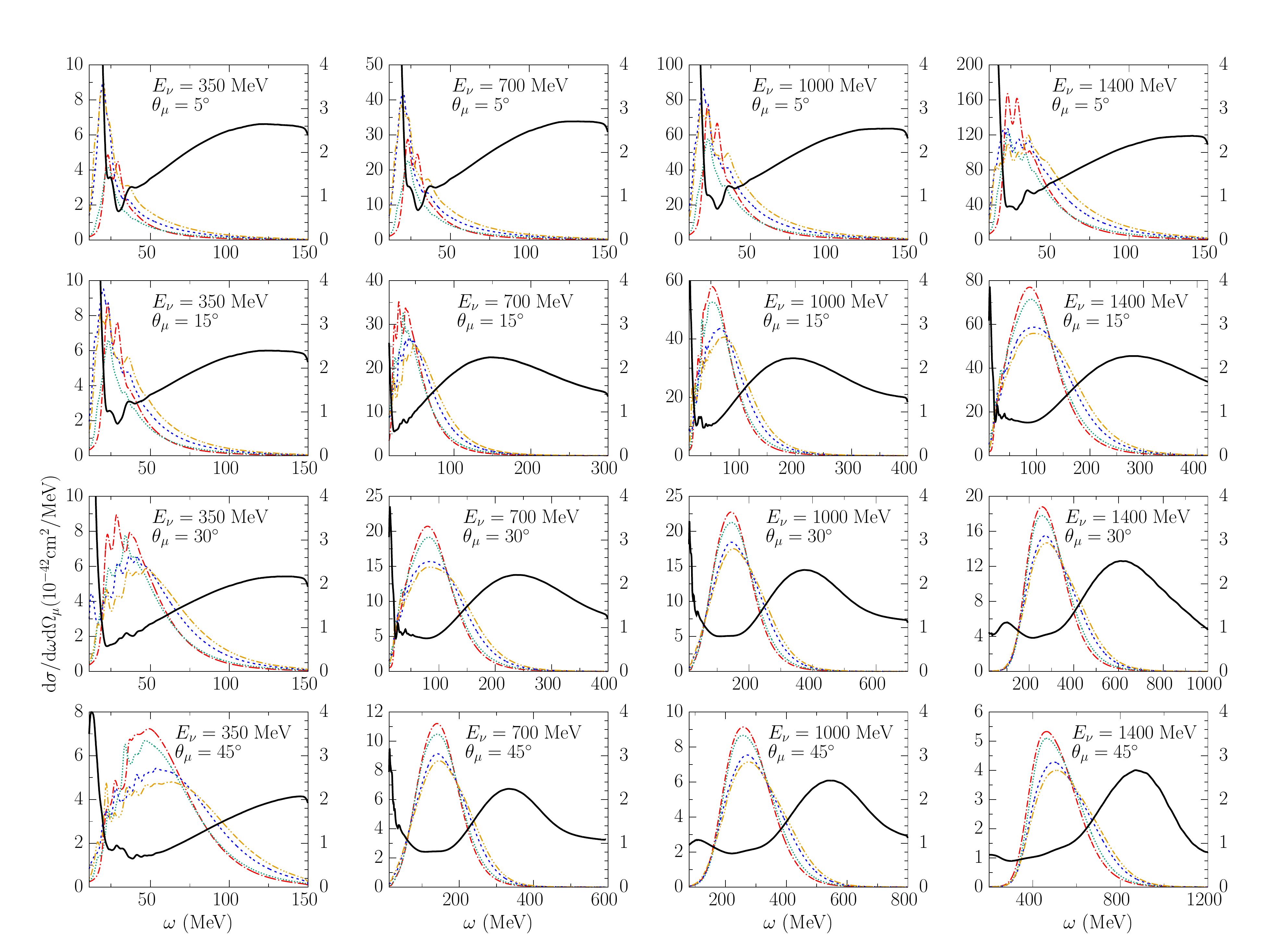}
   \caption{Double differential cross sections (per neutron, left vertical axis) for monochromatic neutrinos of energy $E_\nu$ and muon scattering angle $\theta_\mu$, as a function of $\omega$ (energy transfer). Line key: ${^{12}\textnormal{C}}$ dash--dotted, ${^{16}\textnormal{O}}$ dotted, ${^{40}\textnormal{Ar}}$ dashed and ${^{56}\textnormal{Fe}}$ dash--double dotted. In addition, the cross section ratio $\sigma_{Ar}/\sigma_{C}$ is shown in solid black (right vertical axis).}
   \label{fig:2dplots}
\end{figure*}

In Fig.~\ref{fig:2dplots} double differential cross sections for monochromatic neutrinos are presented, for a set of muon scattering angles $\theta_\mu$. These are shown for ${^{12}\textnormal{C}}$, ${^{16}\textnormal{O}}$, ${^{40}\textnormal{Ar}}$ and ${^{56}\textnormal{Fe}}$. The cross sections are expressed as a function of the energy transfer $\omega$. To facilitate a comparison and to appreciate the differences, the cross sections have been divided by the amount of neutrons in the target nucleus. The ratio of ${^{40}\textnormal{Ar}}$ to ${^{12}\textnormal{C}}$ cross sections is also plotted, to show where the model predictions differ the most between nuclei. \\

The cross sections behave similarly for all nuclei, except for low energies and/or forward scattering angles where peaks show up in the cross-section. The prediction of these collective resonances is a key property of the CRPA approach, which HF calculations alone cannot account for. The peaks are different for the nuclei, differing both in position and relative strength. This is due to the inherently different nuclear structure. Focussing especially on the low--energy region (the left panels) in Fig.~\ref{fig:2dplots} where the collective behaviour is most noticable, we can see that for most $\omega$ values, the strength in ${^{40}\textnormal{Ar}}$ is larger (per nucleon) than it is for e.g. ${^{12}\textnormal{C}}$. It should be noted that this is partially caused by a different threshold energy, as the valence shells of heavier nuclei are less bound than those of lighter ones. Fig.~\ref{fig:2dplots} shows that for higher incoming energies and larger scattering angles, the differential cross section is dominated by the QE peak, corresponding to the intuitive picture of a neutrino interacting with just one of the nucleons in the nucleus. The differential cross sections at these kinematics, too, showcase various differences. In particular one observes that the strength in the QE peak is lower for ${^{40}\textnormal{Ar}}$ and ${^{56}\textnormal{Fe}}$ than ${^{12}\textnormal{C}}$ and ${^{16}\textnormal{O}}$. Indeed, the ratio tends to be smaller than one in the $\omega$--region up to the peak. On the other hand, one also notices that the tails in the case of ${^{40}\textnormal{Ar}}$ are considerably more pronounced. The ratio tends to increase beyond the center of the QE peak. \\ 

\begin{figure}
   \centering
   \includegraphics[width=0.95\columnwidth]{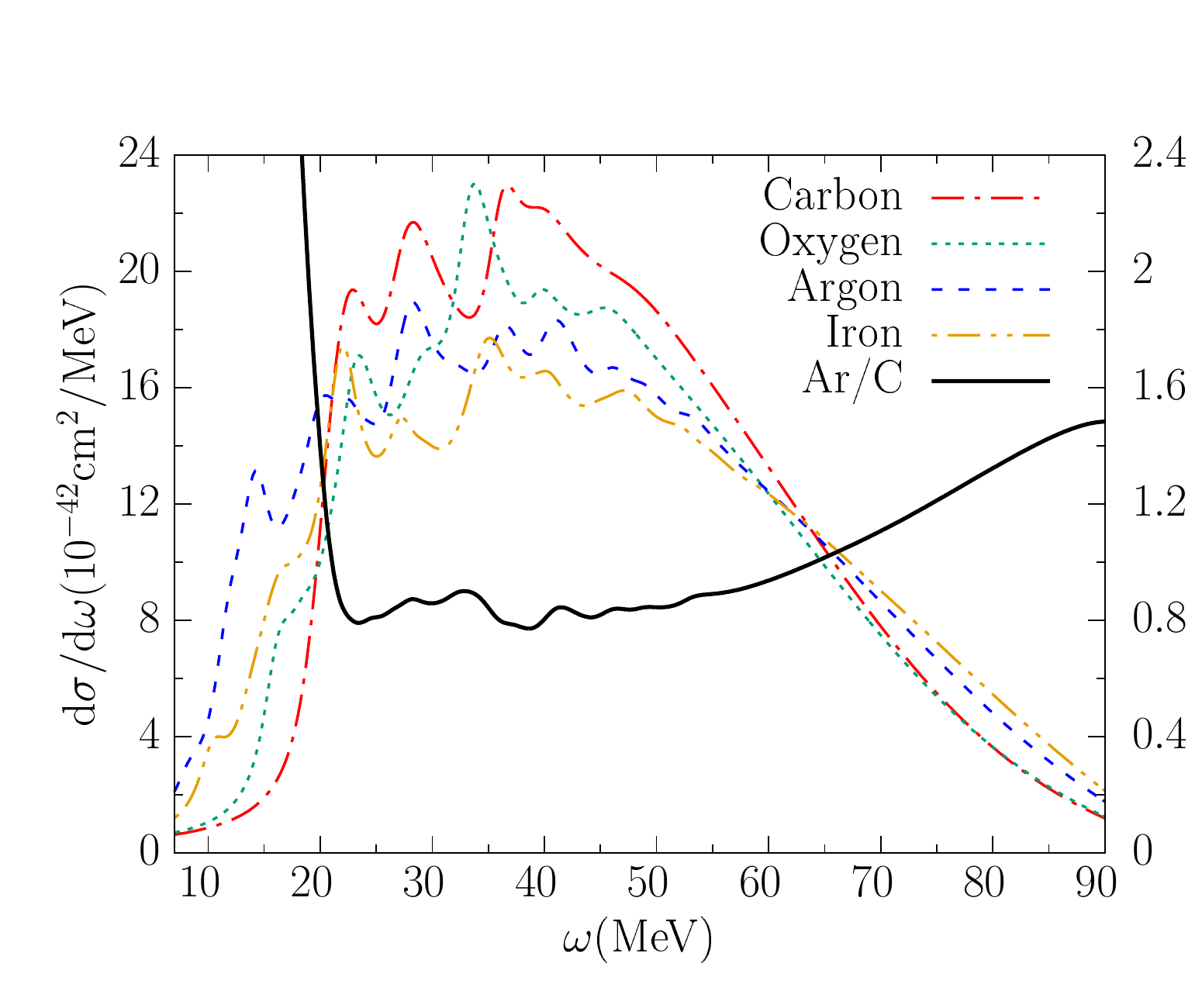}
   \caption{The single differential cross section (per neutron) for CCQE neutrino--nucleus scattering as a function of $\omega$ (left y--axis) for an incoming energy of 200 MeV, along with the ratio between ${^{40}\textnormal{Ar}}$ and ${^{12}\textnormal{C}}$ (right y--axis).}
   \label{fig:singledif}
\end{figure}

In Fig.~\ref{fig:singledif} we take a look at the single differential cross section as a function of $\omega$ for an incoming neutrino energy of 200 MeV. For low--$\omega$ values one sees the inherently different resonance structures causing differences in strength between nuclei. The heavier nuclei dominate low $\omega$ for the same reasons mentioned above. As for double differential cross sections the peak is more pronounced for lighter nuclei. Beyond the peak region, cross sections for heavier nuclei once again start to dominate. The reason for this is twofold. On the one hand, the position of the QE peak shifts to higher $\omega$ values for more backward scattering (see Fig.~\ref{fig:2dplots}). On the other hand, the overall strength at these high $\theta_\mu$ is rather strongly surpressed. The competition between these two effects causes the single differential cross section to be dominated by the contribution of the high--$\omega$ tail of forward processes, where heavy nuclei show a stronger tail. \\

One concludes that, for an ${^{40}\textnormal{Ar}}$ target, modeling the tails of differential cross section is of even more importance than for ${^{12}\textnormal{C}}$. This tail is less accurately described in FG--based models. However, additional effects of the A--dependence when other features such as short--range correlations or meson exchange currents are included are also expected to play an important role at the kinematics for accelerator--based experiments. Due to the fact that 2p2h contributions appear largely in the high--$\omega$ tail of 'QE--like' cross sections \cite{VanCuyck:2016fab,VanCuyck:2017wfn}, we can expect the importance of the tails to become even more marked when taking SRCs into account (we refer to the scaling laws discussed in \cite{Colle:2015ena}). This is subject to future research. \\

\begin{figure}
   \centering
   \includegraphics[width=0.95\columnwidth]{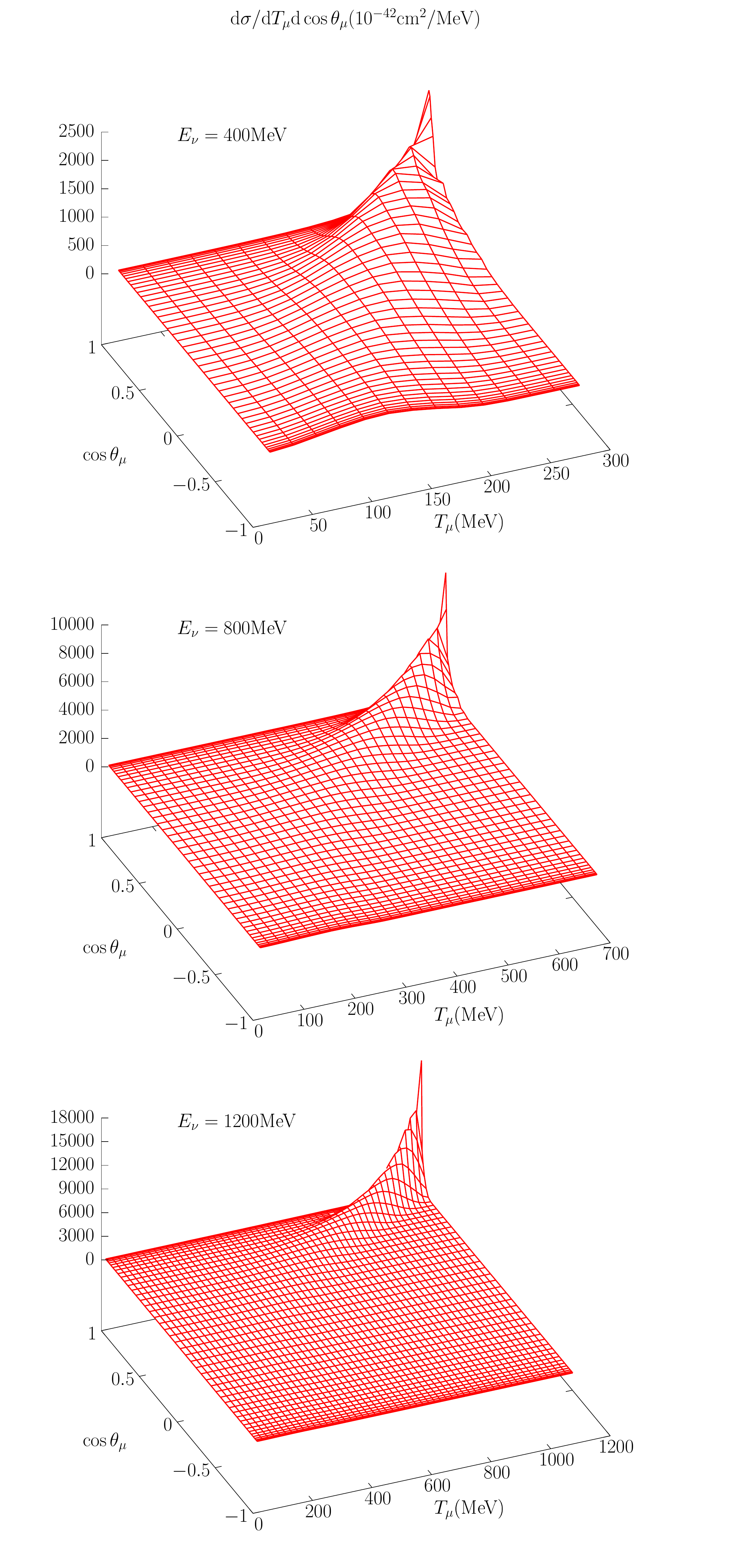}
   \caption{Double differential cross sections for monochromatic neutrinos plotted as a function of the kinetic energy of the outgoing muon $T_\mu$ and the cosine of the outgoing muon angle $\theta_\mu$.} 
   \label{fig:ufx}
\end{figure}

\begin{figure}
   \centering
   \includegraphics[width=0.95\columnwidth]{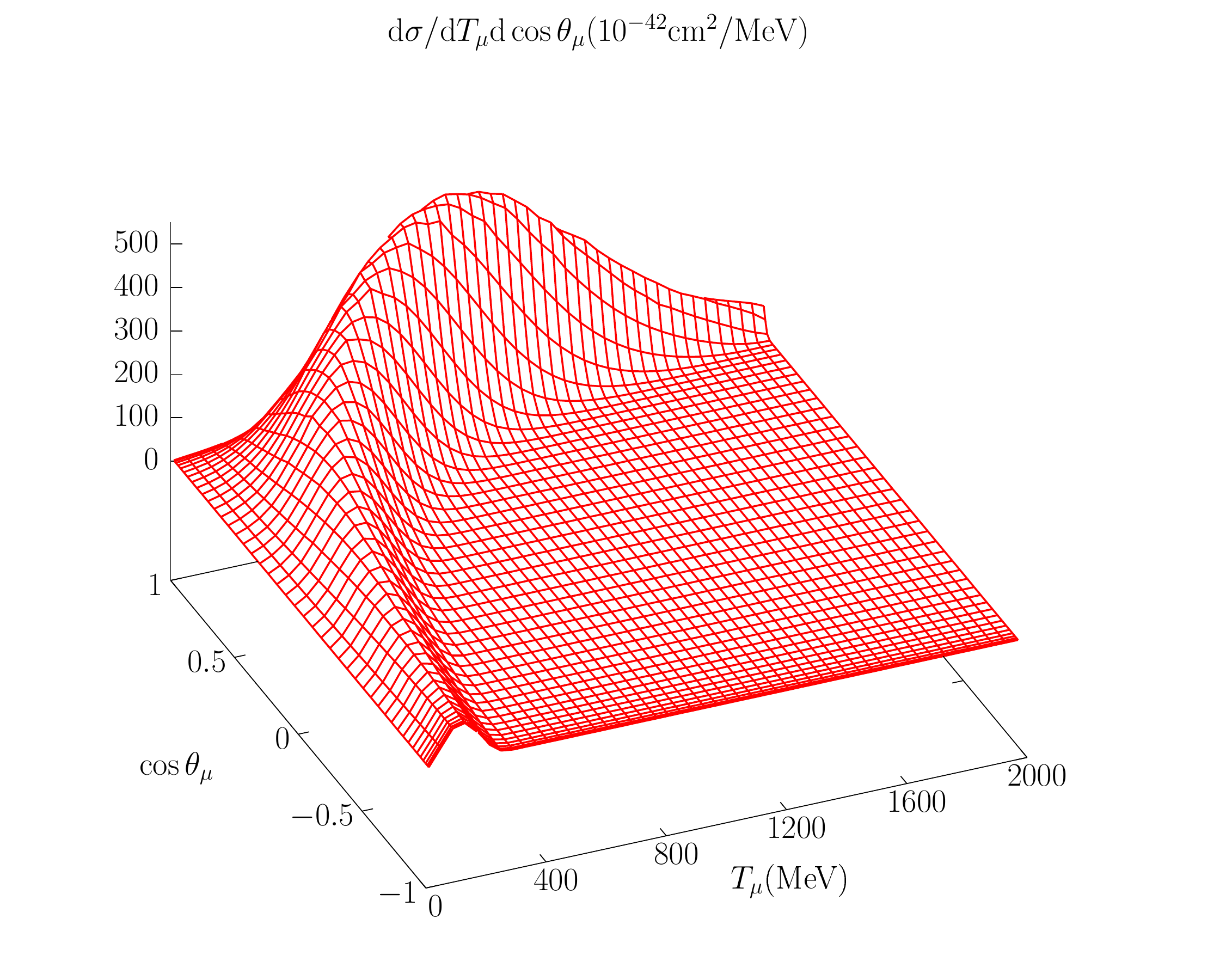}
   \caption{Double differential cross section for CCQE neutrino--argon scattering folded with the MicroBooNE flux. \cite{AguilarArevalo:2008yp}}
   \label{fig:folded}
\end{figure} 

Pictured in Fig.~\ref{fig:ufx} for three incoming neutrino energies are the CCQE neutrino--argon double differential cross sections as a function of $T_\mu$ and $\cos{\theta_\mu}$. The cross sections contain strong contributions for high $T_\mu$ (low $\omega$), and forward scattering angles. For lower energies, one can observe that the reaction strength is spread, with a peak value at high $T_\mu$ (low $\omega$), and forward scattering angles. For higher energies, this spreading is much less noticeable and goes hand in hand with a more pronounced peak value. \\

\begin{figure}
   \centering
   \includegraphics[width=0.60\columnwidth]{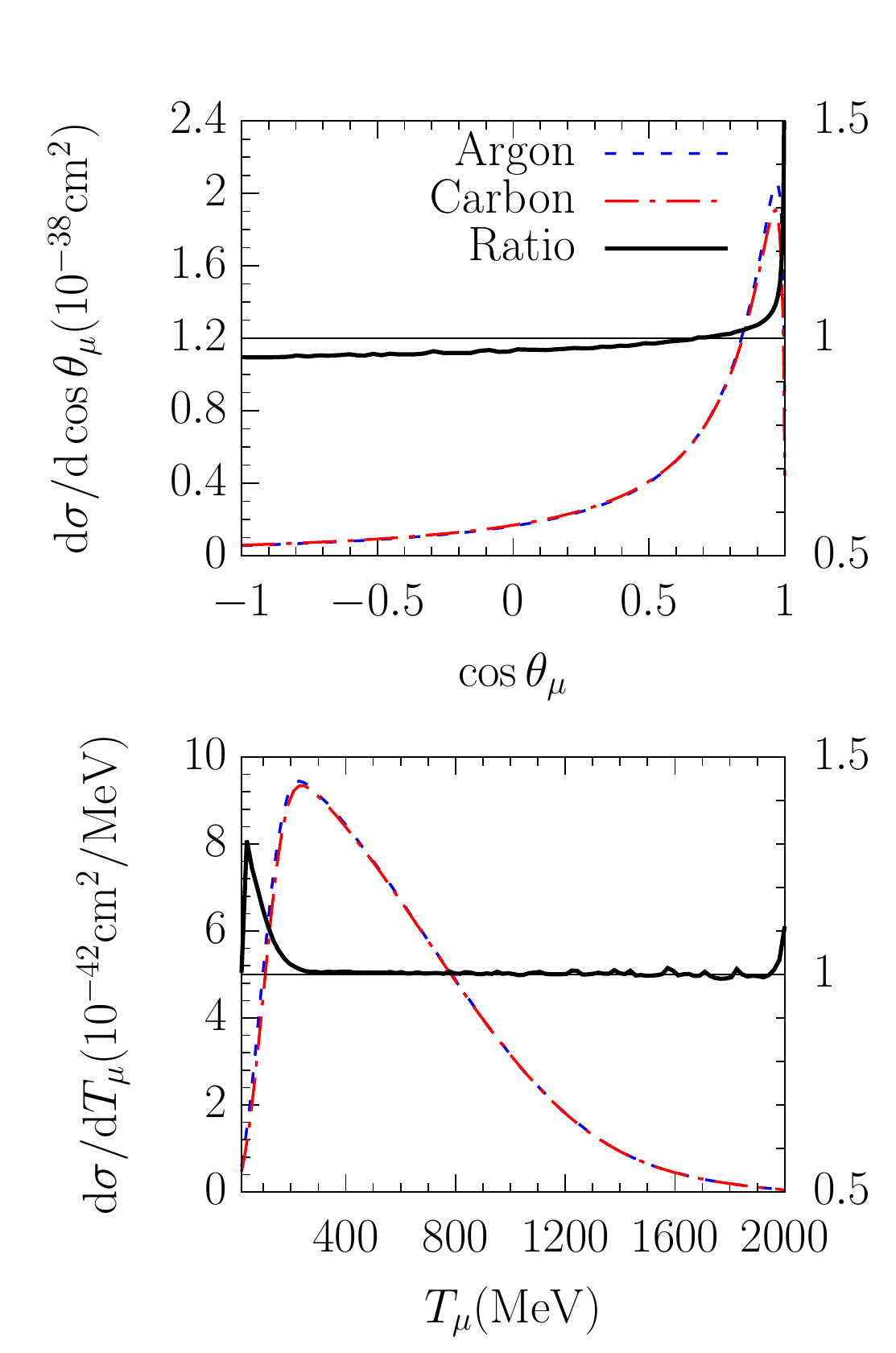}
   \caption{Single differential flux--folded cross sections (per neutron, left vertical axis) as a function of, respectively, the cosine of the outgoing muon angle $\theta_\mu$, and $T_\mu$. Plotted for ${^{12}\textnormal{C}}$ and ${^{40}\textnormal{Ar}}$, in addition to the ratio of the cross sections $\sigma_{Ar}/\sigma_{C}$ (right vertical axis) given by the full line.} 
   \label{fig:neut2}
\end{figure} 
 
Until now, we have been looking at monochromatic neutrinos. In order to compare with data, we need to flux--fold the cross sections. For our purposes, we performed calculations over a range of $E_\nu$ values up to an incoming energy of 2200 MeV, at which point the value of the flux of the Booster Beam line used for MicroBooNE becomes largely negligible \cite{AguilarArevalo:2008yp}. The flux--folded result is displayed in Fig.~\ref{fig:folded}. One can see the importance of forward scattering angles, although, as a result of the folding procedure, the strength is no longer confined to a sharp peak value in the outgoing muon kinetic energy, but smeared out. As could be expected, the higher the kinetic energy of the outgoing muon, the more forward the scattering will be. \\

\begin{figure}
   \centering
   \includegraphics[width=0.95\columnwidth]{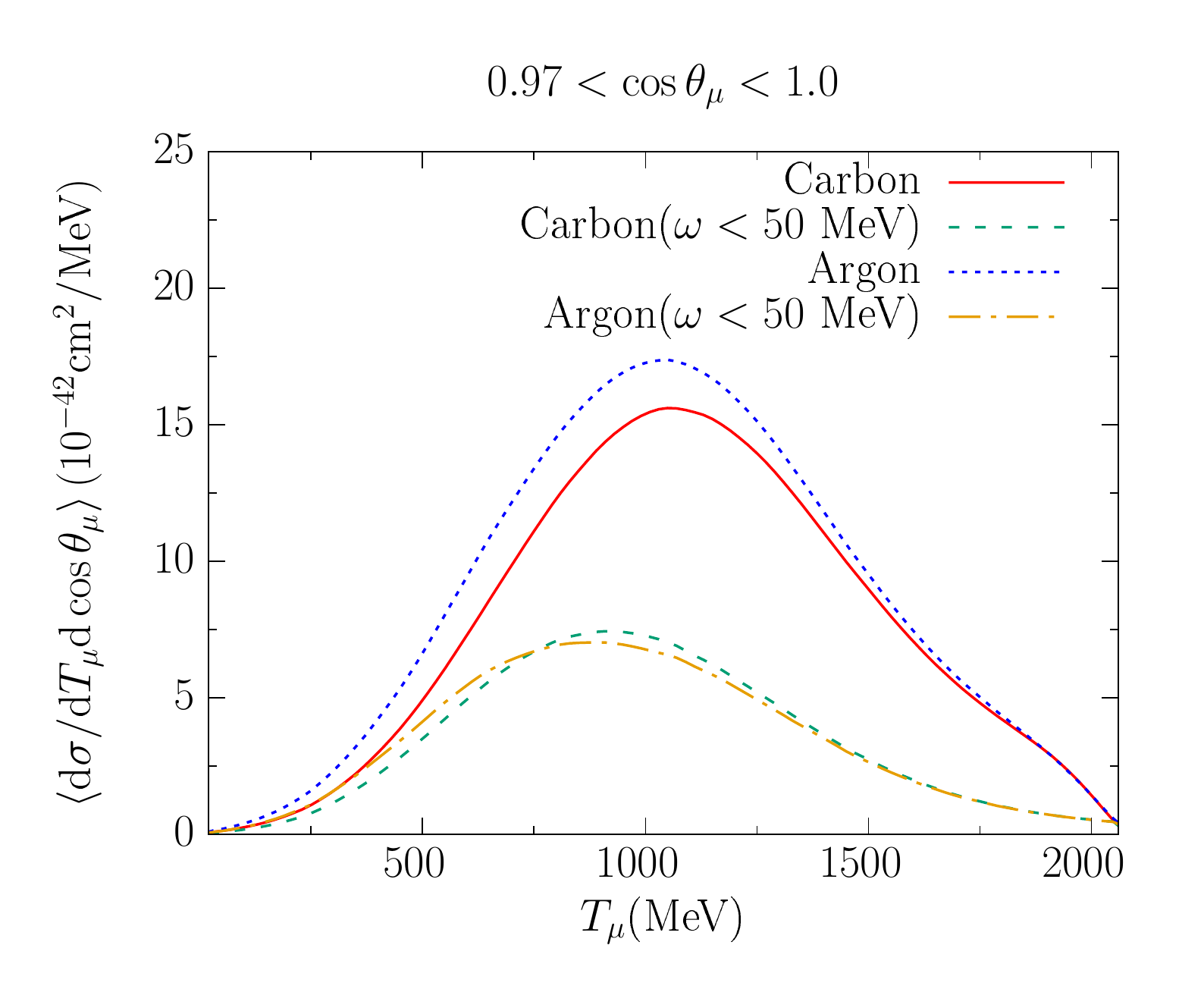}
   \caption{Flux--folded double differential cross section for CCQE neutrino--nucleus scattering, per nucleon. Plotted for ${^{12}\textnormal{C}}$ and ${^{40}\textnormal{Ar}}$, in addition to the $\omega<50$ MeV contribution.}
   \label{fig:lowom}
\end{figure} 

In Fig.~\ref{fig:neut2} we present single differential cross sections off ${^{40}\textnormal{Ar}}$ as a function of $T_\mu$ and $\cos{\theta_\mu}$ and with analogous results for ${^{12}\textnormal{C}}$, along with the ratio. In the first panel, the ratio between the two nuclei starts at around 92\%, increasing slowly. ${^{40}\textnormal{Ar}}$ gets more strength for forward bins with $\cos{\theta} > 0.7$. As one can see in Fig.~\ref{fig:2dplots}, this can be attributed to the much higher contributions for the lowest values of $\omega$, since the threshold energy of ${^{40}\textnormal{Ar}}$ is lower. In the second panel, one can see that except at low values of $T_\mu$, before the peak value of the cross section, the strength per neutron is very similar. The higher strength at low $T_\mu$ is a direct consequence of the high--$\omega$ (low $T_\mu$) tails in Fig.~\ref{fig:2dplots}, which are always higher than those of ${^{12}\textnormal{C}}$ for all incoming energies. Flux--folding will therefore cause an enhanced strength at low $T_\mu$. To further explore the results for forward scattering, we also perform an analysis on the impact of excitations at low $\omega$ for forward scattering angles. In Fig.~\ref{fig:lowom} we display the flux--folded double differential cross section, per neutron and bin averaged for angles $0.97 < \cos{\theta_\mu} < 1.0$. Shown are the results for ${^{12}\textnormal{C}}$ and ${^{40}\textnormal{Ar}}$, as well as the contribution from excitations at values of $\omega$ lower than 50 MeV. In this forward scattering bin, the average reaction strength per active nucleon is higher for ${^{40}\textnormal{Ar}}$ at all values of $T_\mu$. This is in agreement with Fig.~\ref{fig:neut2}, where we see higher reaction strengths for forward scattering. The contribution of the $\omega < 50$ MeV cross section strength is apparent. While it contributes slightly less to the cross section in ${^{40}\textnormal{Ar}}$ relative to ${^{12}\textnormal{C}}$, it is nonetheless still very sizeable, contributing more than 50\% of the strength when $T_\mu$ is lower than $ \approx $ 700 MeV. This is consistent with what is observed in Fig.~\ref{fig:2dplots}, and is likely caused through a combination of nuclear effects, such as the position and strength of the resonances, different single particle energies and occupancies.

\section{Summary}
 We have presented calculations of inclusive CCQE neutrino--nucleus scattering within a Continuum Random Phase Approximation model, to investigate how cross sections compare for various nuclei. \\
 
 In our results we compared ${^{12}\textnormal{C}}$, ${^{16}\textnormal{O}}$, ${^{40}\textnormal{Ar}}$ and ${^{56}\textnormal{Fe}}$ for mono--energetic cross sections to see how model predictions vary. The low--lying nuclear resonances are inherently different. At low $E_\nu$ and also for very forward scattering we observe that ${^{40}\textnormal{Ar}}$ gets more strength than lighter nuclei. Secondly, the quasielastic peak is broader but lower per nucleon for ${^{40}\textnormal{Ar}}$ and ${^{56}\textnormal{Fe}}$ than it is for ${^{12}\textnormal{C}}$ or ${^{16}\textnormal{O}}$. One finds that the reactions strength shifts from the peak to the high--$\omega$ tails. \\
 
We also presented single and double differential inclusive CCQE neutrino--argon scattering cross sections, folded with the flux profile of the Booster Beam line at MicroBooNE. We then took a careful look at the contribution of the reaction strength at values of $\omega$ lower than 50 MeV for forward scattering ($0.97 < \cos{\theta_\mu} < 1.0$). The role of low--energetic excitations is also very important for ${^{40}\textnormal{Ar}}$, contributing considerably to the reaction strength. Additional effects beyond purely QE physics will need to be studied in the future. \\
 
These two aspects of having, on the one hand, increased reaction strength in the high $\omega$ tails and the important role of low--energy excitations at forward scattering angles accentuate the need for detailed theoretical calculations of $\nu-{^{40}\textnormal{Ar}}$ cross sections as we head forward into a future where LArTPCs will play a central role in neutrino detectors. In light of the capability of LArTPCs to perform exclusive measurements containing information on the outgoing hadrons, we will expand our model towards modeling hadronic final states as well in the near future. \\

 \begin{acknowledgments}
   This work was supported by the Research Foundation Flanders (FWO-Flanders) and the Interuniversity Attraction Poles Programme P7/12 initiated by the Belgian Science Policy Office. V.P. acknowledges the support by the National Science Foundation under grant no. PHY--1352106. The computational resources (Stevin Supercomputer Infrastructure) and services used in this work were provided by the VSC (Flemish Supercomputer Center), funded by Ghent University, FWO and the Flemish Government – department EWI.
 \end{acknowledgments}

 \bibliography{biblio}
 \end{document}